\documentstyle[pra,aps,psfig,epsfig]{revtex}
\begin{document}
\draft
\title{PROTON POLARIZABILITY CONTRIBUTION TO THE\\
HYPERFINE SPLITTING IN MUONIC HYDROGEN}
\author{E.V.Cherednikova}
\address{443011, Samara, Pavlov 1, Samara State University,
Russian Federation}
\author{R.N.Faustov\footnote{E-mail:
faustov@theory.sinp.msu.ru}}
\address{117333, Moscow, Vavilov, 40, Scientific Council "Cybernetics" RAS }
\author{A.P.Martynenko\footnote{E-mail:
mart@info.ssu.samara.ru}}
\address{443011, Samara, Pavlov, 1, Department of Theoretical Physics,
Samara State University}

\date{\today}

\maketitle

\begin{abstract}
The contribution of the proton polarizability to the
ground state hyperfine splitting in muonic hydrogen
is evaluated on the basis of modern experimental and
theoretical results on the proton polarized structure functions.
The value of this correction is equal to $4.6(8)\cdot 10^{-4}$
times the Fermi splitting.
\end{abstract}



\immediate\write16{<<WARNING: LINEDRAW macros work with emTeX-dvivers
                    and other drivers supporting emTeX \special's
                    (dviscr, dvihplj, dvidot, dvips, dviwin, etc.) >>}
\newdimen\Lengthunit       \Lengthunit  = 1.5cm
\newcount\Nhalfperiods     \Nhalfperiods= 9
\newcount\magnitude        \magnitude = 1000

\catcode`\*=11
\newdimen\L*   \newdimen\d*   \newdimen\d**
\newdimen\dm*  \newdimen\dd*  \newdimen\dt*
\newdimen\a*   \newdimen\b*   \newdimen\c*
\newdimen\a**  \newdimen\b**
\newdimen\xL*  \newdimen\yL*
\newdimen\rx*  \newdimen\ry*
\newdimen\tmp* \newdimen\linwid*

\newcount\k*   \newcount\l*   \newcount\m*
\newcount\k**  \newcount\l**  \newcount\m**
\newcount\n*   \newcount\dn*  \newcount\r*
\newcount\N*   \newcount\*one \newcount\*two  \*one=1 \*two=2
\newcount\*ths \*ths=1000
\newcount\angle*  \newcount\q*  \newcount\q**
\newcount\angle** \angle**=0
\newcount\sc*     \sc*=0

\newtoks\cos*  \cos*={1}
\newtoks\sin*  \sin*={0}

\catcode`\[=13

\def\rotate(#1){\advance\angle**#1\angle*=\angle**
\q**=\angle*\ifnum\q**<0\q**=-\q**\fi
\ifnum\q**>360\q*=\angle*\divide\q*360\multiply\q*360\advance\angle*-\q*\fi
\ifnum\angle*<0\advance\angle*360\fi\q**=\angle*\divide\q**90\q**=\q**
\def\sgcos*{+}\def\sgsin*{+}\relax
\ifcase\q**\or
 \def\sgcos*{-}\def\sgsin*{+}\or
 \def\sgcos*{-}\def\sgsin*{-}\or
 \def\sgcos*{+}\def\sgsin*{-}\else\fi
\q*=\q**
\multiply\q*90\advance\angle*-\q*
\ifnum\angle*>45\sc*=1\angle*=-\angle*\advance\angle*90\else\sc*=0\fi
\def[##1,##2]{\ifnum\sc*=0\relax
\edef\cs*{\sgcos*.##1}\edef\sn*{\sgsin*.##2}\ifcase\q**\or
 \edef\cs*{\sgcos*.##2}\edef\sn*{\sgsin*.##1}\or
 \edef\cs*{\sgcos*.##1}\edef\sn*{\sgsin*.##2}\or
 \edef\cs*{\sgcos*.##2}\edef\sn*{\sgsin*.##1}\else\fi\else
\edef\cs*{\sgcos*.##2}\edef\sn*{\sgsin*.##1}\ifcase\q**\or
 \edef\cs*{\sgcos*.##1}\edef\sn*{\sgsin*.##2}\or
 \edef\cs*{\sgcos*.##2}\edef\sn*{\sgsin*.##1}\or
 \edef\cs*{\sgcos*.##1}\edef\sn*{\sgsin*.##2}\else\fi\fi
\cos*={\cs*}\sin*={\sn*}\global\edef\gcos*{\cs*}\global\edef\gsin*{\sn*}}\relax
\ifcase\angle*[9999,0]\or
[999,017]\or[999,034]\or[998,052]\or[997,069]\or[996,087]\or
[994,104]\or[992,121]\or[990,139]\or[987,156]\or[984,173]\or
[981,190]\or[978,207]\or[974,224]\or[970,241]\or[965,258]\or
[961,275]\or[956,292]\or[951,309]\or[945,325]\or[939,342]\or
[933,358]\or[927,374]\or[920,390]\or[913,406]\or[906,422]\or
[898,438]\or[891,453]\or[882,469]\or[874,484]\or[866,499]\or
[857,515]\or[848,529]\or[838,544]\or[829,559]\or[819,573]\or
[809,587]\or[798,601]\or[788,615]\or[777,629]\or[766,642]\or
[754,656]\or[743,669]\or[731,681]\or[719,694]\or[707,707]\or
\else[9999,0]\fi}

\catcode`\[=12

\def\GRAPH(hsize=#1)#2{\hbox to #1\Lengthunit{#2\hss}}

\def\Linewidth#1{\global\linwid*=#1\relax
\global\divide\linwid*10\global\multiply\linwid*\mag
\global\divide\linwid*100\special{em:linewidth \the\linwid*}}

\Linewidth{.4pt}
\def\sm*{\special{em:moveto}}
\def\sl*{\special{em:lineto}}
\let\moveto=\sm*
\let\lineto=\sl*
\newbox\spm*   \newbox\spl*
\setbox\spm*\hbox{\sm*}
\setbox\spl*\hbox{\sl*}

\def\mov#1(#2,#3)#4{\rlap{\L*=#1\Lengthunit
\xL*=#2\L* \yL*=#3\L*
\xL*=\xscale\xL* \yL*=\yscale\yL*
\rx* \the\cos*\xL* \tmp* \the\sin*\yL* \advance\rx*-\tmp*
\ry* \the\cos*\yL* \tmp* \the\sin*\xL* \advance\ry*\tmp*
\kern\rx*\raise\ry*\hbox{#4}}}

\def\rmov*(#1,#2)#3{\rlap{\xL*=#1\yL*=#2\relax
\rx* \the\cos*\xL* \tmp* \the\sin*\yL* \advance\rx*-\tmp*
\ry* \the\cos*\yL* \tmp* \the\sin*\xL* \advance\ry*\tmp*
\kern\rx*\raise\ry*\hbox{#3}}}

\def\lin#1(#2,#3){\rlap{\sm*\mov#1(#2,#3){\sl*}}}

\def\arr*(#1,#2,#3){\rmov*(#1\dd*,#1\dt*){\sm*
\rmov*(#2\dd*,#2\dt*){\rmov*(#3\dt*,-#3\dd*){\sl*}}\sm*
\rmov*(#2\dd*,#2\dt*){\rmov*(-#3\dt*,#3\dd*){\sl*}}}}

\def\arrow#1(#2,#3){\rlap{\lin#1(#2,#3)\mov#1(#2,#3){\relax
\d**=-.012\Lengthunit\dd*=#2\d**\dt*=#3\d**
\arr*(1,10,4)\arr*(3,8,4)\arr*(4.8,4.2,3)}}}

\def\arrlin#1(#2,#3){\rlap{\L*=#1\Lengthunit\L*=.5\L*
\lin#1(#2,#3)\rmov*(#2\L*,#3\L*){\arrow.1(#2,#3)}}}

\def\dasharrow#1(#2,#3){\rlap{{\Lengthunit=0.9\Lengthunit
\dashlin#1(#2,#3)\mov#1(#2,#3){\sm*}}\mov#1(#2,#3){\sl*
\d**=-.012\Lengthunit\dd*=#2\d**\dt*=#3\d**
\arr*(1,10,4)\arr*(3,8,4)\arr*(4.8,4.2,3)}}}

\def\clap#1{\hbox to 0pt{\hss #1\hss}}

\def\ind(#1,#2)#3{\rlap{\L*=.1\Lengthunit
\xL*=#1\L* \yL*=#2\L*
\rx* \the\cos*\xL* \tmp* \the\sin*\yL* \advance\rx*-\tmp*
\ry* \the\cos*\yL* \tmp* \the\sin*\xL* \advance\ry*\tmp*
\kern\rx*\raise\ry*\hbox{\lower2pt\clap{$#3$}}}}

\def\sh*(#1,#2)#3{\rlap{\dm*=\the\n*\d**
\xL*=\xscale\dm* \yL*=\yscale\dm* \xL*=#1\xL* \yL*=#2\yL*
\rx* \the\cos*\xL* \tmp* \the\sin*\yL* \advance\rx*-\tmp*
\ry* \the\cos*\yL* \tmp* \the\sin*\xL* \advance\ry*\tmp*
\kern\rx*\raise\ry*\hbox{#3}}}

\def\calcnum*#1(#2,#3){\a*=1000sp\b*=1000sp\a*=#2\a*\b*=#3\b*
\ifdim\a*<0pt\a*-\a*\fi\ifdim\b*<0pt\b*-\b*\fi
\ifdim\a*>\b*\c*=.96\a*\advance\c*.4\b*
\else\c*=.96\b*\advance\c*.4\a*\fi
\k*\a*\multiply\k*\k*\l*\b*\multiply\l*\l*
\m*\k*\advance\m*\l*\n*\c*\r*\n*\multiply\n*\n*
\dn*\m*\advance\dn*-\n*\divide\dn*2\divide\dn*\r*
\advance\r*\dn*
\c*=\the\Nhalfperiods5sp\c*=#1\c*\ifdim\c*<0pt\c*-\c*\fi
\multiply\c*\r*\N*\c*\divide\N*10000}

\def\dashlin#1(#2,#3){\rlap{\calcnum*#1(#2,#3)\relax
\d**=#1\Lengthunit\ifdim\d**<0pt\d**-\d**\fi
\divide\N*2\multiply\N*2\advance\N*\*one
\divide\d**\N*\sm*\n*\*one\sh*(#2,#3){\sl*}\loop
\advance\n*\*one\sh*(#2,#3){\sm*}\advance\n*\*one
\sh*(#2,#3){\sl*}\ifnum\n*<\N*\repeat}}

\def\dashdotlin#1(#2,#3){\rlap{\calcnum*#1(#2,#3)\relax
\d**=#1\Lengthunit\ifdim\d**<0pt\d**-\d**\fi
\divide\N*2\multiply\N*2\advance\N*1\multiply\N*2\relax
\divide\d**\N*\sm*\n*\*two\sh*(#2,#3){\sl*}\loop
\advance\n*\*one\sh*(#2,#3){\kern-1.48pt\lower.5pt\hbox{\rm.}}\relax
\advance\n*\*one\sh*(#2,#3){\sm*}\advance\n*\*two
\sh*(#2,#3){\sl*}\ifnum\n*<\N*\repeat}}

\def\shl*(#1,#2)#3{\kern#1#3\lower#2#3\hbox{\unhcopy\spl*}}

\def\trianglin#1(#2,#3){\rlap{\toks0={#2}\toks1={#3}\calcnum*#1(#2,#3)\relax
\dd*=.57\Lengthunit\dd*=#1\dd*\divide\dd*\N*
\divide\dd*\*ths \multiply\dd*\magnitude
\d**=#1\Lengthunit\ifdim\d**<0pt\d**-\d**\fi
\multiply\N*2\divide\d**\N*\sm*\n*\*one\loop
\shl**{\dd*}\dd*-\dd*\advance\n*2\relax
\ifnum\n*<\N*\repeat\n*\N*\shl**{0pt}}}

\def\wavelin#1(#2,#3){\rlap{\toks0={#2}\toks1={#3}\calcnum*#1(#2,#3)\relax
\dd*=.23\Lengthunit\dd*=#1\dd*\divide\dd*\N*
\divide\dd*\*ths \multiply\dd*\magnitude
\d**=#1\Lengthunit\ifdim\d**<0pt\d**-\d**\fi
\multiply\N*4\divide\d**\N*\sm*\n*\*one\loop
\shl**{\dd*}\dt*=1.3\dd*\advance\n*\*one
\shl**{\dt*}\advance\n*\*one
\shl**{\dd*}\advance\n*\*two
\dd*-\dd*\ifnum\n*<\N*\repeat\n*\N*\shl**{0pt}}}

\def\w*lin(#1,#2){\rlap{\toks0={#1}\toks1={#2}\d**=\Lengthunit\dd*=-.12\d**
\divide\dd*\*ths \multiply\dd*\magnitude
\N*8\divide\d**\N*\sm*\n*\*one\loop
\shl**{\dd*}\dt*=1.3\dd*\advance\n*\*one
\shl**{\dt*}\advance\n*\*one
\shl**{\dd*}\advance\n*\*one
\shl**{0pt}\dd*-\dd*\advance\n*1\ifnum\n*<\N*\repeat}}

\def\l*arc(#1,#2)[#3][#4]{\rlap{\toks0={#1}\toks1={#2}\d**=\Lengthunit
\dd*=#3.037\d**\dd*=#4\dd*\dt*=#3.049\d**\dt*=#4\dt*\ifdim\d**>10mm\relax
\d**=.25\d**\n*\*one\shl**{-\dd*}\n*\*two\shl**{-\dt*}\n*3\relax
\shl**{-\dd*}\n*4\relax\shl**{0pt}\else
\ifdim\d**>5mm\d**=.5\d**\n*\*one\shl**{-\dt*}\n*\*two
\shl**{0pt}\else\n*\*one\shl**{0pt}\fi\fi}}

\def\d*arc(#1,#2)[#3][#4]{\rlap{\toks0={#1}\toks1={#2}\d**=\Lengthunit
\dd*=#3.037\d**\dd*=#4\dd*\d**=.25\d**\sm*\n*\*one\shl**{-\dd*}\relax
\n*3\relax\sh*(#1,#2){\xL*=\xscale\dd*\yL*=\yscale\dd*
\kern#2\xL*\lower#1\yL*\hbox{\sm*}}\n*4\relax\shl**{0pt}}}

\def\shl**#1{\c*=\the\n*\d**\d*=#1\relax
\a*=\the\toks0\c*\b*=\the\toks1\d*\advance\a*-\b*
\b*=\the\toks1\c*\d*=\the\toks0\d*\advance\b*\d*
\a*=\xscale\a*\b*=\yscale\b*
\rx* \the\cos*\a* \tmp* \the\sin*\b* \advance\rx*-\tmp*
\ry* \the\cos*\b* \tmp* \the\sin*\a* \advance\ry*\tmp*
\raise\ry*\rlap{\kern\rx*\unhcopy\spl*}}

\def\wlin*#1(#2,#3)[#4]{\rlap{\toks0={#2}\toks1={#3}\relax
\c*=#1\l*\c*\c*=.01\Lengthunit\m*\c*\divide\l*\m*
\c*=\the\Nhalfperiods5sp\multiply\c*\l*\N*\c*\divide\N*\*ths
\divide\N*2\multiply\N*2\advance\N*\*one
\dd*=.002\Lengthunit\dd*=#4\dd*\multiply\dd*\l*\divide\dd*\N*
\divide\dd*\*ths \multiply\dd*\magnitude
\d**=#1\multiply\N*4\divide\d**\N*\sm*\n*\*one\loop
\shl**{\dd*}\dt*=1.3\dd*\advance\n*\*one
\shl**{\dt*}\advance\n*\*one
\shl**{\dd*}\advance\n*\*two
\dd*-\dd*\ifnum\n*<\N*\repeat\n*\N*\shl**{0pt}}}

\def\wavebox#1{\setbox0\hbox{#1}\relax
\a*=\wd0\advance\a*14pt\b*=\ht0\advance\b*\dp0\advance\b*14pt\relax
\hbox{\kern9pt\relax
\rmov*(0pt,\ht0){\rmov*(-7pt,7pt){\wlin*\a*(1,0)[+]\wlin*\b*(0,-1)[-]}}\relax
\rmov*(\wd0,-\dp0){\rmov*(7pt,-7pt){\wlin*\a*(-1,0)[+]\wlin*\b*(0,1)[-]}}\relax
\box0\kern9pt}}

\def\rectangle#1(#2,#3){\relax
\lin#1(#2,0)\lin#1(0,#3)\mov#1(0,#3){\lin#1(#2,0)}\mov#1(#2,0){\lin#1(0,#3)}}

\def\dashrectangle#1(#2,#3){\dashlin#1(#2,0)\dashlin#1(0,#3)\relax
\mov#1(0,#3){\dashlin#1(#2,0)}\mov#1(#2,0){\dashlin#1(0,#3)}}

\def\waverectangle#1(#2,#3){\L*=#1\Lengthunit\a*=#2\L*\b*=#3\L*
\ifdim\a*<0pt\a*-\a*\def\x*{-1}\else\def\x*{1}\fi
\ifdim\b*<0pt\b*-\b*\def\y*{-1}\else\def\y*{1}\fi
\wlin*\a*(\x*,0)[-]\wlin*\b*(0,\y*)[+]\relax
\mov#1(0,#3){\wlin*\a*(\x*,0)[+]}\mov#1(#2,0){\wlin*\b*(0,\y*)[-]}}

\def\calcparab*{\ifnum\n*>\m*\k*\N*\advance\k*-\n*\else\k*\n*\fi
\a*=\the\k* sp\a*=10\a*\b*\dm*\advance\b*-\a*\k*\b*
\a*=\the\*ths\b*\divide\a*\l*\multiply\a*\k*
\divide\a*\l*\k*\*ths\r*\a*\advance\k*-\r*\dt*=\the\k*\L*}

\def\arcto#1(#2,#3)[#4]{\rlap{\toks0={#2}\toks1={#3}\calcnum*#1(#2,#3)\relax
\dm*=135sp\dm*=#1\dm*\d**=#1\Lengthunit\ifdim\dm*<0pt\dm*-\dm*\fi
\multiply\dm*\r*\a*=.3\dm*\a*=#4\a*\ifdim\a*<0pt\a*-\a*\fi
\advance\dm*\a*\N*\dm*\divide\N*10000\relax
\divide\N*2\multiply\N*2\advance\N*\*one
\L*=-.25\d**\L*=#4\L*\divide\d**\N*\divide\L*\*ths
\m*\N*\divide\m*2\dm*=\the\m*5sp\l*\dm*\sm*\n*\*one\loop
\calcparab*\shl**{-\dt*}\advance\n*1\ifnum\n*<\N*\repeat}}

\def\arrarcto#1(#2,#3)[#4]{\L*=#1\Lengthunit\L*=.54\L*
\arcto#1(#2,#3)[#4]\rmov*(#2\L*,#3\L*){\d*=.457\L*\d*=#4\d*\d**-\d*
\rmov*(#3\d**,#2\d*){\arrow.02(#2,#3)}}}

\def\dasharcto#1(#2,#3)[#4]{\rlap{\toks0={#2}\toks1={#3}\relax
\calcnum*#1(#2,#3)\dm*=\the\N*5sp\a*=.3\dm*\a*=#4\a*\ifdim\a*<0pt\a*-\a*\fi
\advance\dm*\a*\N*\dm*
\divide\N*20\multiply\N*2\advance\N*1\d**=#1\Lengthunit
\L*=-.25\d**\L*=#4\L*\divide\d**\N*\divide\L*\*ths
\m*\N*\divide\m*2\dm*=\the\m*5sp\l*\dm*
\sm*\n*\*one\loop\calcparab*
\shl**{-\dt*}\advance\n*1\ifnum\n*>\N*\else\calcparab*
\sh*(#2,#3){\xL*=#3\dt* \yL*=#2\dt*
\rx* \the\cos*\xL* \tmp* \the\sin*\yL* \advance\rx*\tmp*
\ry* \the\cos*\yL* \tmp* \the\sin*\xL* \advance\ry*-\tmp*
\kern\rx*\lower\ry*\hbox{\sm*}}\fi
\advance\n*1\ifnum\n*<\N*\repeat}}

\def\*shl*#1{\c*=\the\n*\d**\advance\c*#1\a**\d*\dt*\advance\d*#1\b**
\a*=\the\toks0\c*\b*=\the\toks1\d*\advance\a*-\b*
\b*=\the\toks1\c*\d*=\the\toks0\d*\advance\b*\d*
\rx* \the\cos*\a* \tmp* \the\sin*\b* \advance\rx*-\tmp*
\ry* \the\cos*\b* \tmp* \the\sin*\a* \advance\ry*\tmp*
\raise\ry*\rlap{\kern\rx*\unhcopy\spl*}}

\def\calcnormal*#1{\b**=10000sp\a**\b**\k*\n*\advance\k*-\m*
\multiply\a**\k*\divide\a**\m*\a**=#1\a**\ifdim\a**<0pt\a**-\a**\fi
\ifdim\a**>\b**\d*=.96\a**\advance\d*.4\b**
\else\d*=.96\b**\advance\d*.4\a**\fi
\d*=.01\d*\r*\d*\divide\a**\r*\divide\b**\r*
\ifnum\k*<0\a**-\a**\fi\d*=#1\d*\ifdim\d*<0pt\b**-\b**\fi
\k*\a**\a**=\the\k*\dd*\k*\b**\b**=\the\k*\dd*}

\def\wavearcto#1(#2,#3)[#4]{\rlap{\toks0={#2}\toks1={#3}\relax
\calcnum*#1(#2,#3)\c*=\the\N*5sp\a*=.4\c*\a*=#4\a*\ifdim\a*<0pt\a*-\a*\fi
\advance\c*\a*\N*\c*\divide\N*20\multiply\N*2\advance\N*-1\multiply\N*4\relax
\d**=#1\Lengthunit\dd*=.012\d**
\divide\dd*\*ths \multiply\dd*\magnitude
\ifdim\d**<0pt\d**-\d**\fi\L*=.25\d**
\divide\d**\N*\divide\dd*\N*\L*=#4\L*\divide\L*\*ths
\m*\N*\divide\m*2\dm*=\the\m*0sp\l*\dm*
\sm*\n*\*one\loop\calcnormal*{#4}\calcparab*
\*shl*{1}\advance\n*\*one\calcparab*
\*shl*{1.3}\advance\n*\*one\calcparab*
\*shl*{1}\advance\n*2\dd*-\dd*\ifnum\n*<\N*\repeat\n*\N*\shl**{0pt}}}

\def\triangarcto#1(#2,#3)[#4]{\rlap{\toks0={#2}\toks1={#3}\relax
\calcnum*#1(#2,#3)\c*=\the\N*5sp\a*=.4\c*\a*=#4\a*\ifdim\a*<0pt\a*-\a*\fi
\advance\c*\a*\N*\c*\divide\N*20\multiply\N*2\advance\N*-1\multiply\N*2\relax
\d**=#1\Lengthunit\dd*=.012\d**
\divide\dd*\*ths \multiply\dd*\magnitude
\ifdim\d**<0pt\d**-\d**\fi\L*=.25\d**
\divide\d**\N*\divide\dd*\N*\L*=#4\L*\divide\L*\*ths
\m*\N*\divide\m*2\dm*=\the\m*0sp\l*\dm*
\sm*\n*\*one\loop\calcnormal*{#4}\calcparab*
\*shl*{1}\advance\n*2\dd*-\dd*\ifnum\n*<\N*\repeat\n*\N*\shl**{0pt}}}

\def\hr*#1{\L*=\xscale\Lengthunit\ifnum
\angle**=0\clap{\vrule width#1\L* height.1pt}\else
\L*=#1\L*\L*=.5\L*\rmov*(-\L*,0pt){\sm*}\rmov*(\L*,0pt){\sl*}\fi}

\def\shade#1[#2]{\rlap{\Lengthunit=#1\Lengthunit
\special{em:linewidth .001pt}\relax
\mov(0,#2.05){\hr*{.994}}\mov(0,#2.1){\hr*{.980}}\relax
\mov(0,#2.15){\hr*{.953}}\mov(0,#2.2){\hr*{.916}}\relax
\mov(0,#2.25){\hr*{.867}}\mov(0,#2.3){\hr*{.798}}\relax
\mov(0,#2.35){\hr*{.715}}\mov(0,#2.4){\hr*{.603}}\relax
\mov(0,#2.45){\hr*{.435}}\special{em:linewidth \the\linwid*}}}

\def\dshade#1[#2]{\rlap{\special{em:linewidth .001pt}\relax
\Lengthunit=#1\Lengthunit\if#2-\def\t*{+}\else\def\t*{-}\fi
\mov(0,\t*.025){\relax
\mov(0,#2.05){\hr*{.995}}\mov(0,#2.1){\hr*{.988}}\relax
\mov(0,#2.15){\hr*{.969}}\mov(0,#2.2){\hr*{.937}}\relax
\mov(0,#2.25){\hr*{.893}}\mov(0,#2.3){\hr*{.836}}\relax
\mov(0,#2.35){\hr*{.760}}\mov(0,#2.4){\hr*{.662}}\relax
\mov(0,#2.45){\hr*{.531}}\mov(0,#2.5){\hr*{.320}}\relax
\special{em:linewidth \the\linwid*}}}}

\def\vdot{\rlap{\kern-1.9pt\lower1.8pt\hbox{$\scriptstyle\bullet$}}}
\def\vtimes{\rlap{\kern-3pt\lower1.8pt\hbox{$\scriptstyle\times$}}}
\def\vDot{\rlap{\kern-2.3pt\lower2.7pt\hbox{$\bullet$}}}
\def\vTimes{\rlap{\kern-3.6pt\lower2.4pt\hbox{$\times$}}}

\def\arc(#1)[#2,#3]{{\k*=#2\l*=#3\m*=\l*
\advance\m*-6\ifnum\k*>\l*\relax\else
{\rotate(#2)\mov(#1,0){\sm*}}\loop
\ifnum\k*<\m*\advance\k*5{\rotate(\k*)\mov(#1,0){\sl*}}\repeat
{\rotate(#3)\mov(#1,0){\sl*}}\fi}}

\def\dasharc(#1)[#2,#3]{{\k**=#2\n*=#3\advance\n*-1\advance\n*-\k**
\L*=1000sp\L*#1\L* \multiply\L*\n* \multiply\L*\Nhalfperiods
\divide\L*57\N*\L* \divide\N*2000\ifnum\N*=0\N*1\fi
\r*\n*  \divide\r*\N* \ifnum\r*<2\r*2\fi
\m**\r* \divide\m**2 \l**\r* \advance\l**-\m** \N*\n* \divide\N*\r*
\k**\r* \multiply\k**\N* \dn*\n* \advance\dn*-\k** \divide\dn*2\advance\dn*\*one
\r*\l** \divide\r*2\advance\dn*\r* \advance\N*-2\k**#2\relax
\ifnum\l**<6{\rotate(#2)\mov(#1,0){\sm*}}\advance\k**\dn*
{\rotate(\k**)\mov(#1,0){\sl*}}\advance\k**\m**
{\rotate(\k**)\mov(#1,0){\sm*}}\loop
\advance\k**\l**{\rotate(\k**)\mov(#1,0){\sl*}}\advance\k**\m**
{\rotate(\k**)\mov(#1,0){\sm*}}\advance\N*-1\ifnum\N*>0\repeat
{\rotate(#3)\mov(#1,0){\sl*}}\else\advance\k**\dn*
\arc(#1)[#2,\k**]\loop\advance\k**\m** \r*\k**
\advance\k**\l** {\arc(#1)[\r*,\k**]}\relax
\advance\N*-1\ifnum\N*>0\repeat
\advance\k**\m**\arc(#1)[\k**,#3]\fi}}

\def\triangarc#1(#2)[#3,#4]{{\k**=#3\n*=#4\advance\n*-\k**
\L*=1000sp\L*#2\L* \multiply\L*\n* \multiply\L*\Nhalfperiods
\divide\L*57\N*\L* \divide\N*1000\ifnum\N*=0\N*1\fi
\d**=#2\Lengthunit \d*\d** \divide\d*57\multiply\d*\n*
\r*\n*  \divide\r*\N* \ifnum\r*<2\r*2\fi
\m**\r* \divide\m**2 \l**\r* \advance\l**-\m** \N*\n* \divide\N*\r*
\dt*\d* \divide\dt*\N* \dt*.5\dt* \dt*#1\dt*
\divide\dt*1000\multiply\dt*\magnitude
\k**\r* \multiply\k**\N* \dn*\n* \advance\dn*-\k** \divide\dn*2\relax
\r*\l** \divide\r*2\advance\dn*\r* \advance\N*-1\k**#3\relax
{\rotate(#3)\mov(#2,0){\sm*}}\advance\k**\dn*
{\rotate(\k**)\mov(#2,0){\sl*}}\advance\k**-\m**\advance\l**\m**\loop\dt*-\dt*
\d*\d** \advance\d*\dt*
\advance\k**\l**{\rotate(\k**)\rmov*(\d*,0pt){\sl*}}%
\advance\N*-1\ifnum\N*>0\repeat\advance\k**\m**
{\rotate(\k**)\mov(#2,0){\sl*}}{\rotate(#4)\mov(#2,0){\sl*}}}}

\def\wavearc#1(#2)[#3,#4]{{\k**=#3\n*=#4\advance\n*-\k**
\L*=4000sp\L*#2\L* \multiply\L*\n* \multiply\L*\Nhalfperiods
\divide\L*57\N*\L* \divide\N*1000\ifnum\N*=0\N*1\fi
\d**=#2\Lengthunit \d*\d** \divide\d*57\multiply\d*\n*
\r*\n*  \divide\r*\N* \ifnum\r*=0\r*1\fi
\m**\r* \divide\m**2 \l**\r* \advance\l**-\m** \N*\n* \divide\N*\r*
\dt*\d* \divide\dt*\N* \dt*.7\dt* \dt*#1\dt*
\divide\dt*1000\multiply\dt*\magnitude
\k**\r* \multiply\k**\N* \dn*\n* \advance\dn*-\k** \divide\dn*2\relax
\divide\N*4\advance\N*-1\k**#3\relax
{\rotate(#3)\mov(#2,0){\sm*}}\advance\k**\dn*
{\rotate(\k**)\mov(#2,0){\sl*}}\advance\k**-\m**\advance\l**\m**\loop\dt*-\dt*
\d*\d** \advance\d*\dt* \dd*\d** \advance\dd*1.3\dt*
\advance\k**\r*{\rotate(\k**)\rmov*(\d*,0pt){\sl*}}\relax
\advance\k**\r*{\rotate(\k**)\rmov*(\dd*,0pt){\sl*}}\relax
\advance\k**\r*{\rotate(\k**)\rmov*(\d*,0pt){\sl*}}\relax
\advance\k**\r*
\advance\N*-1\ifnum\N*>0\repeat\advance\k**\m**
{\rotate(\k**)\mov(#2,0){\sl*}}{\rotate(#4)\mov(#2,0){\sl*}}}}

\def\gmov*#1(#2,#3)#4{\rlap{\L*=#1\Lengthunit
\xL*=#2\L* \yL*=#3\L*
\rx* \gcos*\xL* \tmp* \gsin*\yL* \advance\rx*-\tmp*
\ry* \gcos*\yL* \tmp* \gsin*\xL* \advance\ry*\tmp*
\rx*=\xscale\rx* \ry*=\yscale\ry*
\xL* \the\cos*\rx* \tmp* \the\sin*\ry* \advance\xL*-\tmp*
\yL* \the\cos*\ry* \tmp* \the\sin*\rx* \advance\yL*\tmp*
\kern\xL*\raise\yL*\hbox{#4}}}

\def\rgmov*(#1,#2)#3{\rlap{\xL*#1\yL*#2\relax
\rx* \gcos*\xL* \tmp* \gsin*\yL* \advance\rx*-\tmp*
\ry* \gcos*\yL* \tmp* \gsin*\xL* \advance\ry*\tmp*
\rx*=\xscale\rx* \ry*=\yscale\ry*
\xL* \the\cos*\rx* \tmp* \the\sin*\ry* \advance\xL*-\tmp*
\yL* \the\cos*\ry* \tmp* \the\sin*\rx* \advance\yL*\tmp*
\kern\xL*\raise\yL*\hbox{#3}}}

\def\Earc(#1)[#2,#3][#4,#5]{{\k*=#2\l*=#3\m*=\l*
\advance\m*-6\ifnum\k*>\l*\relax\else\def\xscale{#4}\def\yscale{#5}\relax
{\angle**0\rotate(#2)}\gmov*(#1,0){\sm*}\loop
\ifnum\k*<\m*\advance\k*5\relax
{\angle**0\rotate(\k*)}\gmov*(#1,0){\sl*}\repeat
{\angle**0\rotate(#3)}\gmov*(#1,0){\sl*}\relax
\def\xscale{1}\def\yscale{1}\fi}}

\def\dashEarc(#1)[#2,#3][#4,#5]{{\k**=#2\n*=#3\advance\n*-1\advance\n*-\k**
\L*=1000sp\L*#1\L* \multiply\L*\n* \multiply\L*\Nhalfperiods
\divide\L*57\N*\L* \divide\N*2000\ifnum\N*=0\N*1\fi
\r*\n*  \divide\r*\N* \ifnum\r*<2\r*2\fi
\m**\r* \divide\m**2 \l**\r* \advance\l**-\m** \N*\n* \divide\N*\r*
\k**\r*\multiply\k**\N* \dn*\n* \advance\dn*-\k** \divide\dn*2\advance\dn*\*one
\r*\l** \divide\r*2\advance\dn*\r* \advance\N*-2\k**#2\relax
\ifnum\l**<6\def\xscale{#4}\def\yscale{#5}\relax
{\angle**0\rotate(#2)}\gmov*(#1,0){\sm*}\advance\k**\dn*
{\angle**0\rotate(\k**)}\gmov*(#1,0){\sl*}\advance\k**\m**
{\angle**0\rotate(\k**)}\gmov*(#1,0){\sm*}\loop
\advance\k**\l**{\angle**0\rotate(\k**)}\gmov*(#1,0){\sl*}\advance\k**\m**
{\angle**0\rotate(\k**)}\gmov*(#1,0){\sm*}\advance\N*-1\ifnum\N*>0\repeat
{\angle**0\rotate(#3)}\gmov*(#1,0){\sl*}\def\xscale{1}\def\yscale{1}\else
\advance\k**\dn* \Earc(#1)[#2,\k**][#4,#5]\loop\advance\k**\m** \r*\k**
\advance\k**\l** {\Earc(#1)[\r*,\k**][#4,#5]}\relax
\advance\N*-1\ifnum\N*>0\repeat
\advance\k**\m**\Earc(#1)[\k**,#3][#4,#5]\fi}}

\def\triangEarc#1(#2)[#3,#4][#5,#6]{{\k**=#3\n*=#4\advance\n*-\k**
\L*=1000sp\L*#2\L* \multiply\L*\n* \multiply\L*\Nhalfperiods
\divide\L*57\N*\L* \divide\N*1000\ifnum\N*=0\N*1\fi
\d**=#2\Lengthunit \d*\d** \divide\d*57\multiply\d*\n*
\r*\n*  \divide\r*\N* \ifnum\r*<2\r*2\fi
\m**\r* \divide\m**2 \l**\r* \advance\l**-\m** \N*\n* \divide\N*\r*
\dt*\d* \divide\dt*\N* \dt*.5\dt* \dt*#1\dt*
\divide\dt*1000\multiply\dt*\magnitude
\k**\r* \multiply\k**\N* \dn*\n* \advance\dn*-\k** \divide\dn*2\relax
\r*\l** \divide\r*2\advance\dn*\r* \advance\N*-1\k**#3\relax
\def\xscale{#5}\def\yscale{#6}\relax
{\angle**0\rotate(#3)}\gmov*(#2,0){\sm*}\advance\k**\dn*
{\angle**0\rotate(\k**)}\gmov*(#2,0){\sl*}\advance\k**-\m**
\advance\l**\m**\loop\dt*-\dt* \d*\d** \advance\d*\dt*
\advance\k**\l**{\angle**0\rotate(\k**)}\rgmov*(\d*,0pt){\sl*}\relax
\advance\N*-1\ifnum\N*>0\repeat\advance\k**\m**
{\angle**0\rotate(\k**)}\gmov*(#2,0){\sl*}\relax
{\angle**0\rotate(#4)}\gmov*(#2,0){\sl*}\def\xscale{1}\def\yscale{1}}}

\def\waveEarc#1(#2)[#3,#4][#5,#6]{{\k**=#3\n*=#4\advance\n*-\k**
\L*=4000sp\L*#2\L* \multiply\L*\n* \multiply\L*\Nhalfperiods
\divide\L*57\N*\L* \divide\N*1000\ifnum\N*=0\N*1\fi
\d**=#2\Lengthunit \d*\d** \divide\d*57\multiply\d*\n*
\r*\n*  \divide\r*\N* \ifnum\r*=0\r*1\fi
\m**\r* \divide\m**2 \l**\r* \advance\l**-\m** \N*\n* \divide\N*\r*
\dt*\d* \divide\dt*\N* \dt*.7\dt* \dt*#1\dt*
\divide\dt*1000\multiply\dt*\magnitude
\k**\r* \multiply\k**\N* \dn*\n* \advance\dn*-\k** \divide\dn*2\relax
\divide\N*4\advance\N*-1\k**#3\def\xscale{#5}\def\yscale{#6}\relax
{\angle**0\rotate(#3)}\gmov*(#2,0){\sm*}\advance\k**\dn*
{\angle**0\rotate(\k**)}\gmov*(#2,0){\sl*}\advance\k**-\m**
\advance\l**\m**\loop\dt*-\dt*
\d*\d** \advance\d*\dt* \dd*\d** \advance\dd*1.3\dt*
\advance\k**\r*{\angle**0\rotate(\k**)}\rgmov*(\d*,0pt){\sl*}\relax
\advance\k**\r*{\angle**0\rotate(\k**)}\rgmov*(\dd*,0pt){\sl*}\relax
\advance\k**\r*{\angle**0\rotate(\k**)}\rgmov*(\d*,0pt){\sl*}\relax
\advance\k**\r*
\advance\N*-1\ifnum\N*>0\repeat\advance\k**\m**
{\angle**0\rotate(\k**)}\gmov*(#2,0){\sl*}\relax
{\angle**0\rotate(#4)}\gmov*(#2,0){\sl*}\def\xscale{1}\def\yscale{1}}}

\newcount\CatcodeOfAtSign
\CatcodeOfAtSign=\the\catcode`\@
\catcode`\@=11
\def\@arc#1[#2][#3]{\rlap{\Lengthunit=#1\Lengthunit
\sm*\l*arc(#2.1914,#3.0381)[#2][#3]\relax
\mov(#2.1914,#3.0381){\l*arc(#2.1622,#3.1084)[#2][#3]}\relax
\mov(#2.3536,#3.1465){\l*arc(#2.1084,#3.1622)[#2][#3]}\relax
\mov(#2.4619,#3.3086){\l*arc(#2.0381,#3.1914)[#2][#3]}}}

\def\dash@arc#1[#2][#3]{\rlap{\Lengthunit=#1\Lengthunit
\d*arc(#2.1914,#3.0381)[#2][#3]\relax
\mov(#2.1914,#3.0381){\d*arc(#2.1622,#3.1084)[#2][#3]}\relax
\mov(#2.3536,#3.1465){\d*arc(#2.1084,#3.1622)[#2][#3]}\relax
\mov(#2.4619,#3.3086){\d*arc(#2.0381,#3.1914)[#2][#3]}}}

\def\wave@arc#1[#2][#3]{\rlap{\Lengthunit=#1\Lengthunit
\w*lin(#2.1914,#3.0381)\relax
\mov(#2.1914,#3.0381){\w*lin(#2.1622,#3.1084)}\relax
\mov(#2.3536,#3.1465){\w*lin(#2.1084,#3.1622)}\relax
\mov(#2.4619,#3.3086){\w*lin(#2.0381,#3.1914)}}}

\def\bezier#1(#2,#3)(#4,#5)(#6,#7){\N*#1\l*\N* \advance\l*\*one
\d* #4\Lengthunit \advance\d* -#2\Lengthunit \multiply\d* \*two
\b* #6\Lengthunit \advance\b* -#2\Lengthunit
\advance\b*-\d* \divide\b*\N*
\d** #5\Lengthunit \advance\d** -#3\Lengthunit \multiply\d** \*two
\b** #7\Lengthunit \advance\b** -#3\Lengthunit
\advance\b** -\d** \divide\b**\N*
\mov(#2,#3){\sm*{\loop\ifnum\m*<\l*
\a*\m*\b* \advance\a*\d* \divide\a*\N* \multiply\a*\m*
\a**\m*\b** \advance\a**\d** \divide\a**\N* \multiply\a**\m*
\rmov*(\a*,\a**){\unhcopy\spl*}\advance\m*\*one\repeat}}}

\catcode`\*=12

\newcount\n@ast
\def\n@ast@#1{\n@ast0\relax\get@ast@#1\end}
\def\get@ast@#1{\ifx#1\end\let\next\relax\else
\ifx#1*\advance\n@ast1\fi\let\next\get@ast@\fi\next}

\newif\if@up \newif\if@dwn
\def\up@down@#1{\@upfalse\@dwnfalse
\if#1u\@uptrue\fi\if#1U\@uptrue\fi\if#1+\@uptrue\fi
\if#1d\@dwntrue\fi\if#1D\@dwntrue\fi\if#1-\@dwntrue\fi}

\def\halfcirc#1(#2)[#3]{{\Lengthunit=#2\Lengthunit\up@down@{#3}\relax
\if@up\mov(0,.5){\@arc[-][-]\@arc[+][-]}\fi
\if@dwn\mov(0,-.5){\@arc[-][+]\@arc[+][+]}\fi
\def\lft{\mov(0,.5){\@arc[-][-]}\mov(0,-.5){\@arc[-][+]}}\relax
\def\rght{\mov(0,.5){\@arc[+][-]}\mov(0,-.5){\@arc[+][+]}}\relax
\if#3l\lft\fi\if#3L\lft\fi\if#3r\rght\fi\if#3R\rght\fi
\n@ast@{#1}\relax
\ifnum\n@ast>0\if@up\shade[+]\fi\if@dwn\shade[-]\fi\fi
\ifnum\n@ast>1\if@up\dshade[+]\fi\if@dwn\dshade[-]\fi\fi}}

\def\halfdashcirc(#1)[#2]{{\Lengthunit=#1\Lengthunit\up@down@{#2}\relax
\if@up\mov(0,.5){\dash@arc[-][-]\dash@arc[+][-]}\fi
\if@dwn\mov(0,-.5){\dash@arc[-][+]\dash@arc[+][+]}\fi
\def\lft{\mov(0,.5){\dash@arc[-][-]}\mov(0,-.5){\dash@arc[-][+]}}\relax
\def\rght{\mov(0,.5){\dash@arc[+][-]}\mov(0,-.5){\dash@arc[+][+]}}\relax
\if#2l\lft\fi\if#2L\lft\fi\if#2r\rght\fi\if#2R\rght\fi}}

\def\halfwavecirc(#1)[#2]{{\Lengthunit=#1\Lengthunit\up@down@{#2}\relax
\if@up\mov(0,.5){\wave@arc[-][-]\wave@arc[+][-]}\fi
\if@dwn\mov(0,-.5){\wave@arc[-][+]\wave@arc[+][+]}\fi
\def\lft{\mov(0,.5){\wave@arc[-][-]}\mov(0,-.5){\wave@arc[-][+]}}\relax
\def\rght{\mov(0,.5){\wave@arc[+][-]}\mov(0,-.5){\wave@arc[+][+]}}\relax
\if#2l\lft\fi\if#2L\lft\fi\if#2r\rght\fi\if#2R\rght\fi}}

\catcode`\*=11

\def\Circle#1(#2){\halfcirc#1(#2)[u]\halfcirc#1(#2)[d]\n@ast@{#1}\relax
\ifnum\n@ast>0\L*=\xscale\Lengthunit
\ifnum\angle**=0\clap{\vrule width#2\L* height.1pt}\else
\L*=#2\L*\L*=.5\L*\special{em:linewidth .001pt}\relax
\rmov*(-\L*,0pt){\sm*}\rmov*(\L*,0pt){\sl*}\relax
\special{em:linewidth \the\linwid*}\fi\fi}

\catcode`\*=12

\def\wavecirc(#1){\halfwavecirc(#1)[u]\halfwavecirc(#1)[d]}

\def\dashcirc(#1){\halfdashcirc(#1)[u]\halfdashcirc(#1)[d]}

\def\xscale{1}
\def\yscale{1}

\def\Ellipse#1(#2)[#3,#4]{\def\xscale{#3}\def\yscale{#4}\relax
\Circle#1(#2)\def\xscale{1}\def\yscale{1}}

\def\dashEllipse(#1)[#2,#3]{\def\xscale{#2}\def\yscale{#3}\relax
\dashcirc(#1)\def\xscale{1}\def\yscale{1}}

\def\waveEllipse(#1)[#2,#3]{\def\xscale{#2}\def\yscale{#3}\relax
\wavecirc(#1)\def\xscale{1}\def\yscale{1}}

\def\halfEllipse#1(#2)[#3][#4,#5]{\def\xscale{#4}\def\yscale{#5}\relax
\halfcirc#1(#2)[#3]\def\xscale{1}\def\yscale{1}}

\def\halfdashEllipse(#1)[#2][#3,#4]{\def\xscale{#3}\def\yscale{#4}\relax
\halfdashcirc(#1)[#2]\def\xscale{1}\def\yscale{1}}

\def\halfwaveEllipse(#1)[#2][#3,#4]{\def\xscale{#3}\def\yscale{#4}\relax
\halfwavecirc(#1)[#2]\def\xscale{1}\def\yscale{1}}

\catcode`\@=\the\CatcodeOfAtSign

\section{Introduction}

The investigation of the energy spectra of hydrogenic atoms (positronium,
muonium, hydrogen atom, muonic hydrogen, et. al) as well as lepton anomalous
magnetic moments provides a test of quantum electrodynamics
and the theory of electromagnetic bound states with very high accuracy. Moreover,
the values of the fundamental physical constants (the particle masses, fine
structure constant, proton charge radius, etc.) can be determined more precisely.
Inclusion
of new simple atomic systems in the range of the experimental investigations
can lead to the significant progress in solving of these problems. The measurement
of the muonic hydrogen Lamb shift at PSI with a precision of 30 ppm will allow
to improve our knowledge of the proton charge radius by an order of magnitude
\cite{FK}.
Another important problem is connected with the measurement of the ground
state hyperfine splitting (HFS) in muonic hydrogen \cite{BR}.
In the case of
electronic hydrogen HFS was measured with extremely high accuracy many years
ago \cite{Hellwig}:
\begin{equation}
{\rm \Delta \nu^{exp}_{HFS}({\rm e p})=1420405.7517667(9)~~{\rm kHz}.}
\end{equation}

The corresponding theoretical expression of the hydrogen hyperfine
splitting can be written in the form (${\rm \Delta E^{th}_{HFS}=
2\pi\hbar\Delta\nu^{th}_{HFS}}$) \cite{EGS}:
\begin{equation}
{\rm \Delta E^{th}_{HFS}=E_F(1+\delta^{QED}+\delta^S+\delta^P),~~E_F=\frac{8}{3}
\alpha^4\frac{\mu_Pm_1^2m_2^2}{(m_1+m_2)^3},}
\end{equation}
where ${\rm \mu_P}$ is the proton magnetic moment, ${\rm m_1}$, ${\rm m_2}$
are the masses
of the electron (muon) and proton. The calculation of different corrections to
${\rm E_F}$ has a long history. Modern status in the theory of hydrogenic atoms
was presented in details in \cite{EGS}. ${\rm \delta^{QED}}$ denotes the contribution
of higher-order quantumelectrodynamical effects. This correction is known
with an accuracy ${\rm 10^{-7}}$ \cite{EGS}.
Corrections ${\rm \delta^S}$ and
${\rm \delta^P}$ take into account the influence of strong interaction. ${\rm \delta^S}$
describes the effects of the proton finite-size and recoil contribution.
${\rm \delta^P}$ is the correction due to the proton polarizability.
The calculation of the nucleon electromagnetic form factors and the
polarized structure functions on the basis of QCD inspired effective
field theories \cite{DT} with high degree of accuracy can lead to the
determination of the contributions ${\rm \delta^S}$ and ${\rm \delta^P}$
in the hydrogen HFS.
The progress in solving of these problems strongly depends also on the
successful realization of the program for the study of the nucleon excitations
at CEBAF \cite{B}.

The comparison of the theoretical result (2) for electronic hydrogen without
the unknown proton polarizability correction with the experimental value (1)
yields:
\begin{equation}
{\rm \frac{\Delta E_{HFS}^{th}-\Delta
E_{HFS}^{exp}}{E_F}=-4.5(1.1)\cdot 10^{-6},}
\end{equation}
where the main sources of uncertainty in (3) are the inaccuracy of the proton
form factor parameterization (dipole fit etc.) and the contradictory
experimental data on the proton radius. The general structure of the
expression for the muonic hydrogen hyperfine splitting coincides with (2).
The QED contribution emerges after replacement ${\rm m_e\rightarrow m_\mu}$.
The corrections ${\rm \delta^S(\mu p)}$ and ${\rm \delta^P(\mu p)}$ also
depend on the lepton mass. So, there are no simple relations
between ${\rm \delta^S}$ and ${\rm\delta^P}$ for muonic and electronic
hydrogen. But the inclusion of the muonic hydrogen HFS
in the sphere of the theoretical and experimental investigations
can shed additional light on the proton polarizability and proton
structure corrections in both systems.

\section{General formalism}

The main part of the one-loop proton structure correction is
determined by the following expression (Zemach correction) \cite{AZ,BY}:
\begin{equation}
{\rm \Delta E=E_F\frac{2\alpha\mu}{\pi^2}\int \frac{d\vec p}{(p^2+b^2)^2}\Biggl
[\frac{G_E(-\vec p^2)G_M(-\vec p^2)}{1+\kappa}-1\Biggr]
=E_F(-2\mu\alpha)R_p,~b=\alpha\mu,}
\end{equation}
where ${\rm R_p}$ is the Zemach radius, $\mu$ is reduced mass.
In the coordinate representation the Zemach correction is determined
by the magnetic moment density ${\rm \rho_M(r)}$ and by the density
of electric charge ${\rm \rho_E(r)}$.
The value of ${\rm R_p}$ can be
obtained in the analytical form by using the dipole parameterization
for the proton electromagnetic form factors ${\rm G_E}$ and ${\rm G_M}$:
\begin{equation}
{\rm R_p}=\frac{35}{8\Lambda}-\frac{12b}{\Lambda^2}+\frac{105
b^2}{4\Lambda^3}+ O\left(\frac{b^3}{\Lambda^4}\right),
\end{equation}
where ${\rm \Lambda=0.898 m_p}$. Correction to the mass independent
part $35/8\Lambda$ in eq. (5) of order ${\rm O(b/\Lambda)}$ is small
due to the first power of the fine structure constant. Relative order
contribution of the Zemach correction to the hydrogen HFS is the
following:
\begin{equation}
{\rm electronic~ hydrogen}:~~~{\rm R_p}=1.025~{\rm fm},~\delta({\rm Zemach})=-38.72~{\rm ppm},
\end{equation}
\begin{equation}
{\rm muonic~ hydrogen}:~~~{\rm R_p}=1.022~{\rm fm},~\delta({\rm
Zemach})= -71.80\cdot 10^{-4},
\end{equation}
Other possible parameterizations for the nucleon electromagnetic form factors
\cite{Simon} lead to changing the Zemach correction by 3-4 $\%$.
The second recoil part of the one-loop contribution to the HFS in hydrogen
is equal to ${\rm 5.22\cdot 10^{-6}E_F}$ \cite{BY}. Other
uncertainty of the expression (2) is connected with the proton polarizability
correction $\delta^P$. This contribution can be obtained from the two-photon
exchange diagrams as illustrated in Fig. 1. The corresponding amplitudes of
the virtual Compton scattering on the
proton can be expressed through nucleon polarized structure
functions $\rm G_1(\nu,Q^2)$ and $\rm G_2(\nu,Q^2)$. The inelastic contribution
of the diagrams (a), (b) Fig. 1 to the hydrogen HFS was studied in \cite{I,SD,V,G,Z},
where the electron mass was neglected. Preserving exact dependence on the
muon mass in the case of muonic hydrogen we can present the proton polarizability
contribution to the HFS in the form:
\begin{equation}
{\rm \Delta E_{HFS}^P=\frac{Z\alpha m_1}{2\pi m_2(1+\kappa)}E_F(\Delta_1+\Delta_2)=
(\delta_1^P+\delta_2^P)E_F=\delta^PE_F,}
\end{equation}
\begin{equation}
{\rm \Delta_1=\int_0^\infty\frac{dQ^2}{Q^2}\left\{\frac{9}{4}F_2^2(Q^2)\beta_0(\sigma)-
4m_2^3\int_{\nu_{th}}^\infty\frac{d\nu}{\nu}\beta_1(\sigma,\theta)
G_1(\nu,Q^2)\right\},}
\end{equation}
\begin{equation}
{\rm \Delta_2=-12m_2^2\int_0^\infty\frac{dQ^2}{Q^2}
\int_{\nu_{th}}^\infty d\nu\beta_2(\sigma,\theta)
G_2(\nu,Q^2),}
\end{equation}
where ${\rm \nu_{th}}$ determines the pion-nucleon threshold:
\begin{equation}
{\rm \nu_{th}=m_\pi+\frac{m_\pi^2+Q^2}{2m_2},}
\end{equation}
and the functions ${\rm \beta_{0,1,2}}$ have the form:
\begin{equation}
{\rm \beta_0(\sigma)=2\frac{\sqrt{1+\sigma}-1}{\sigma},~~~\sigma=\frac{4m^2_1}{Q^2},}
\end{equation}
\begin{equation}
{\rm \beta_1(\sigma,\theta)=\frac{\theta}{\sigma(-1+\sigma\theta)}\Biggl[
\frac{-2\sqrt{1+\sigma}+2-4\sigma^2-2\sigma}{\sqrt{1+\sigma}}+}
\end{equation}
\begin{displaymath}
{\rm +\frac{2\sigma(-\theta^2+\theta^{3/2}\sqrt{1+\theta}+\theta+2)}
{\sqrt{\theta(1+\theta)}}\Biggr],~\theta=\frac{\nu^2}{Q^2}}
\end{displaymath}
\begin{equation}
{\rm \beta_2(\sigma,\theta)=\frac{2}{\sigma(-1+\sigma\theta)}\left[1-\sqrt{1+\sigma}+
\sigma\left(-\theta+\sqrt{\theta(1+\theta)}\right)\right].}
\end{equation}
${\rm F_2(Q^2)}$ is the Pauli form factor of the proton and the
proton anomalous magnetic moment $\kappa$=1.792847386(63)
\cite{RPP}. For past years there were not enough experimental
data and theoretical information about proton spin-dependent
structure functions. So, the previous study of the contribution
${\rm \Delta E_{HFS}^P}$ contains only estimation of the proton
polarizability effects: $\delta^P\sim 1\div 2$ {\rm ppm}, or the
calculation of main resonance contributions \cite{V,G,Z,FMS}. The
theoretical bound for the proton polarizability contribution to
the HFS of electronic hydrogen is $|\delta^P|\le 4~{\rm ppm}$
\cite{Rafael}. New estimation of the contribution $\delta^P$ in
the hydrogen atom was done in \cite{FM2000} on the basis of
modern experimental data and theoretical results on the structure
functions ${\rm G_{1,2}(\nu,Q^2)}$. By virtue of the fact that
the muon to electron mass ratio ${\rm m_\mu/m_e}$=206.768266 the
lepton mass dependent terms in the functions $\beta_i$ give
appreciable contribution to the $\delta^P$. It is our purpose
here to calculate the correction $\delta^P$ for muonic hydrogen
with the account muon mass dependent terms in (8)-(10). Previously
we have considered also the other possible source of small
uncertainty in the HFS interval, connected with the hadronic
vacuum polarization \cite{FM1998}.

The polarized structure functions ${\rm g_1(\nu,Q^2)}$ and ${\rm g_2(\nu,Q^2)}$ enter
in the antisymmetric part of hadronic tensor ${\rm W_{\mu\nu}}$, describing
lepton-nucleon deep inelastic scattering (DIS) \cite{Close}:
\begin{equation}
{\rm W_{\mu\nu}=W_{\mu\nu}^{[S]}+W_{\mu\nu}^{[A]},}
\end{equation}
\begin{equation}
{\rm W_{\mu\nu}^{[S]}=\left(-g_{\mu\nu}+\frac{q_\mu q_\nu}{q^2}\right)W_1(\nu,Q^2)+
\left(P_\mu-\frac{P\cdot q}{q^2}q_\mu\right)\left(P_\nu-\frac{P\cdot q}{q^2}q_\nu\right)
\frac{W_2(\nu,Q^2)}{m_2^2},}
\end{equation}
\begin{equation}
{\rm W_{\mu\nu}^{[A]}=\epsilon_{\mu\nu\alpha\beta}q^\alpha\left\{S^\beta
\frac{g_1(\nu,Q^2)}{P\cdot q}+[(P\cdot q)S^\beta-(S\cdot q)P^\beta]
\frac{g_2(\nu,Q^2)}{(P\cdot q)^2}\right\},}
\end{equation}
${\rm \epsilon_{\mu\nu\alpha\beta}}$ is the totally antisymmetric
tensor in four dimensions, ${\rm g_1(\nu,Q^2)}$ = ${\rm m_2^2\nu G_1(\nu,Q^2)}$,
${\rm g_2(\nu,Q^2)=m_2\nu^2 G_2(\nu,Q^2)}$, P is the four-momentum of
the nucleon, ${\rm x=Q^2/2m_2\nu}$ is the Bjorken variable, S is
the proton spin four-vector, normalized to ${\rm S^2=-1}$, ${\rm
q^2=-Q^2}$ is the square of the four-momentum transfer. The
invariant quantity $P\cdot q$ is related to the energy transfer
$\nu$ in the proton rest frame: ${\rm P\cdot q=m_2\nu}$. The
invariant mass of the electroproduced hadronic system W is then
${\rm W^2=m_2^2+2m_2\nu-Q^2}$ = ${\rm m_2^2+Q^2(1/x-1)}$. Here ${\rm W_1}$
and ${\rm W_2}$ are the structure functions for unpolarized
scattering. In the DIS regime the invariant mass W must be
greater than any resonance in the nucleon. The threshold between
the resonance region and the deep-inelastic region is not well
defined, but it is usually taken to be at about ${\rm W^2=4}~GeV^2$.

\begin{figure*}[t!]
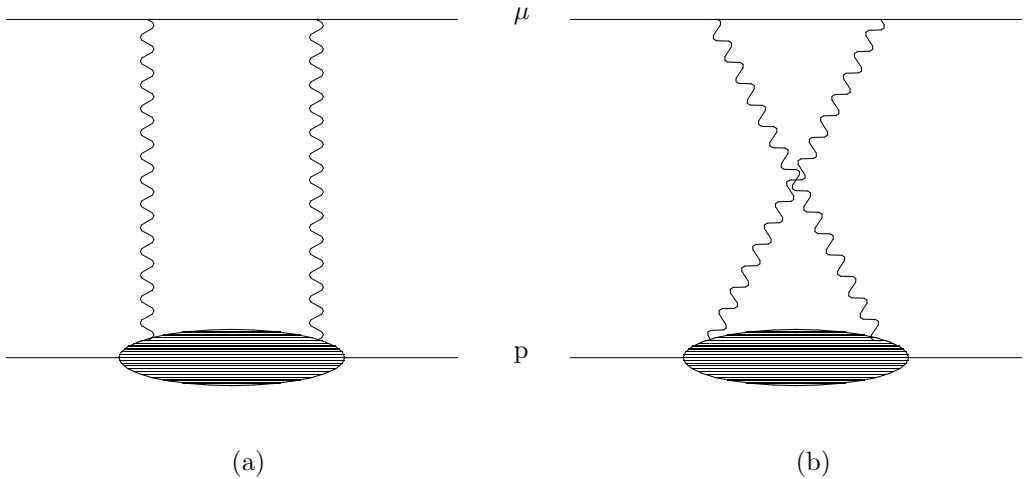

\magnitude=2000
\GRAPH(hsize=15){
\mov(0,0){\lin(1,0)}%
\mov(3,0){\lin(1,0)}%
\mov(5,0){\lin(1,0)}%
\mov(4.5,0){p}%
\mov(4.5,3){$\mu$}%
\mov(8,0){\lin(1,0)}%
\mov(0,3){\lin(4,0)}%
\mov(5,3){\lin(4,0)}%
\mov(1.25,3){\wavelin(0,-2.85)}%
\mov(2.75,3){\wavelin(0,-2.85)}%
\mov(2.,0){\Ellipse*(0.5)[4,1]}%
\mov(7.,0){\Ellipse*(0.5)[4,1]}%
\mov(7.,-1.){(b)}%
\mov(2.,-1.){(a)}%
\mov(7.75,3){\wavelin(-1.5,-2.85)}%
\mov(6.25,3){\wavelin(1.5,-2.85)}%
}
\vspace*{13pt}
\caption{Feynman diagrams for proton polarizability correction
to the hydrogen HFS. }
\end{figure*}

Hadronic tensor ${\rm W_{\mu\nu}}$ is proportional to the
imaginary part of the off-shell Compton amplitude for the forward
scattering of virtual photons on nucleons: ${\rm \gamma^\ast
N\rightarrow \gamma^\ast N}$. The photon-nucleon interaction
depends on the photon polarization as well as on the nucleon one.
This gives four independent helicity amplitudes of the
form ${\rm M_{ab;cd}}$, with a, b, c, d values for the helicities
of the photon and nucleon initial and final states:
\begin{displaymath}
{\rm
M_{1,1/2;1,1/2},~~~M_{1,-1/2;1,-1/2},~~~M_{0,1/2;0,1/2},~~~M_{1,1/2;0,-1/2}.}
\end{displaymath}
These components correspond to the four structure
functions ${\rm W_1, W_2, g_1, g_2}$. All other possible
combinations of initial and final photon and nucleon helicities
are related to the above by time reversal and parity
transformation.

The proton spin structure functions can be measured in the
inelastic scattering of polarized electrons on polarized protons.
Recent improvements in polarized lepton beams and nucleon targets
have made it possible to make accurate measurements of
nucleon polarized structure functions ${\rm g_{1,2}}$ in experiments at
SLAC, CERN and DESY \cite{Abe1,Abe2,Anthony,Mitchell,Adams,Adeva,Hughes}.
The spin dependent structure functions can be expressed in terms of virtual
photon-absorption cross sections \cite{Close}:
\begin{equation}
{\rm g_1(\nu,Q^2)=\frac{m_2\cdot K}{8\pi^2\alpha(1+Q^2/\nu^2)}\left[\sigma_{1/2}
(\nu,Q^2)-\sigma_{3/2}(\nu,Q^2)+\frac{2\sqrt{Q^2}}{\nu}\sigma_{TL}(\nu,Q^2)\right]}
\end{equation}
\begin{equation}
{\rm g_2(\nu,Q^2)=\frac{m_2\cdot K}{8\pi^2\alpha(1+Q^2/\nu^2)}\left[-\sigma_{1/2}
(\nu,Q^2)+\sigma_{3/2}(\nu,Q^2)+\frac{2\nu}{\sqrt{Q^2}}\sigma_{TL}(\nu,Q^2)\right]}
\end{equation}
where ${\rm K=\nu-\frac{Q^2}{2m_2}}$ is the Hand kinematical flux
factor for virtual photons, $\sigma_{1/2}$, $\sigma_{3/2}$ are
the virtual photoabsorption transverse cross sections for the total
photon-nucleon helicity of 1/2 and 3/2 respectively,
${\rm \sigma_{TL}}$ is the interference term between the
transverse and longitudinal photon-nucleon amplitudes. In this
work we calculate contribution ${\rm \Delta E_{HFS}^P}$ on the
basis of the latest experimental data on structure functions ${\rm
g_{1,2}(\nu,Q^2)}$ and theoretical predictions for the cross sections
${\rm \sigma_{1/2,3/2,TL}}$.

\section{Polarized structure functions}

\subsection{Deep-Inelastic region}

Our calculation of the contribution ${\rm \Delta E_{HFS}^P}$ in the DIS region
${\rm (W^2\geq 4~ GeV^2)}$ is based as on the recent
experimental data \cite{Abe1,Abe2,Anthony,Adeva}, so on the evolution
equations for the polarized parton densities. The structure function
${\rm g_1}$ is related to the polarized quark and gluon distributions by
\cite{Altarelli,Kumano}
\begin{equation}
{\rm g_1(x,Q^2)=\frac{<e^2>}{2}\left[C_{NS}\otimes\Delta q_{NS}+C_S\otimes
\Delta\Sigma+2n_fC_g\otimes\Delta g\right],}
\end{equation}
where ${\rm <e^2>=n_f^{-1}\sum_{i=1}^ne_i^2}$, $\otimes$ denotes
convolution with respect to x. The nonsinglet and singlet
quark distributions are defined as:
\begin{equation}
{\rm \Delta q_{NS}=\sum_{i=1}^{n_f}\left(\frac{e_i^2}{<e^2>}-1\right)\left(\Delta
q_i+\Delta\bar q_i\right),}
\end{equation}
where ${\rm \Delta q_i(x,Q^2)=q_+(x,Q^2)-q_-(x,Q^2)}$ measures to what degree the parton
of flavour q "remembers" its parent proton polarization, ${\rm \Delta g(x,Q^2)}$ is
the longitudinally polarized gluon density, probed at a scale $Q^2$. The
evolution equations for the polarized parton densities are given by
\cite{Altarelli,Kumano}
\begin{equation}
{\rm \frac{d}{dt}\Delta q_{NS}=\frac{\alpha_s(t)}{2\pi}P_{qq}^{NS}\otimes\Delta
q_{NS},}
\end{equation}
\begin{displaymath}
{\rm \frac{d}{dt}\left(\Delta\Sigma\atop\Delta g\right)=\frac{\alpha_s(t)}{2\pi}
\left(\begin{array}{cc}
P_{qq}^S & 2 n_f P_{qg}^S \\
P_{gq}^S & P_{gg}^S
\end{array}\right)\otimes\left(\Delta\Sigma\atop\Delta g\right),}
\end{displaymath}
where ${\rm t=ln(Q^2/\Lambda_0^2)}$. The coefficient functions C and the polarized
splitting functions P are now known at NLO. To mitigate possible higher
twist contributions we used the Fortran program for solving the ${\rm Q^2}$ evolution
equations, suggested in \cite{Kumano}, only in the region of ${\rm Q^2\ge 1~Gev^2}$.
The comparison of obtained results for the polarized structure function ${\rm g_1(x,Q^2)}$
with experimental data \cite{Abe1,Abe2,Anthony,Adeva} is presented in Fig.2-4 at
different ${\rm Q^2}$. Recent experimental data \cite{Abe1,Abe2,Anthony,Adeva}
show that there are still large experimental errors for the polarized
structure function ${\rm g_1(x,Q^2)}$. So, application of the evolution equations
in the nonresonance region of ${\rm Q^2\ge 1~GeV^2}$ allows to decrease substantially
the theoretical errors in the $\delta^P$ calculation. In the region
of ${\rm Q^2\le 1~Gev^2}$ we used the experimental data for ${\rm g_1(x,Q^2)}$.
All of the data, including the SMC data at ${\rm Q^2\leq 1 ~GeV^2}$, were fitted
by the parameterization:
\begin{equation}
{\rm g_1(x,Q^2)=a_1 x^{a_2}(1+a_3x+a_4x^2)[1+a_5f(Q^2)]F_1(x,Q^2),}
\end{equation}
where ${\rm F_1=W_1m_2}$. The coefficients of the fits and different models
for the form of the ${\rm Q^2}$ dependence may be found in \cite{Abe1}.
Numerical integration in (8) was performed with the
${\rm f(Q^2)=-\ln Q^2}$ (fit IV), corresponding to the perturbative  QCD behaviour.
We have extrapolated relation (23) to the region near ${\rm Q^2=0}$.
Calculation of the second part of the correction ${\rm \delta^P}$ in (8)
for the nonresonance region was performed by means of the
Wandzura-Wilczek relation between spin structure functions ${\rm
g_1(x,Q^2)}$ and ${\rm g_2(x,Q^2)}$:
\begin{equation}
{\rm g_2^{WW}(x,Q^2)=-g_1(x,Q^2)+\int_x^1
g_1(t,Q^2)\frac{dt}{t},~~~g_2^{WW}\approx g_2.}
\end{equation}
Higher twist terms contribute to ${\rm g_2(x,Q^2)}$ as well, but they
are small enough. So when the transverse asymmetry $A_\perp$ is not
measured or is poorly known, ${\rm g_2=g_2^{WW}}$ is often used.

\begin{figure}[htbp]\vspace*{0.0cm}
\epsfxsize=0.9\textwidth
\centerline{\psfig{figure=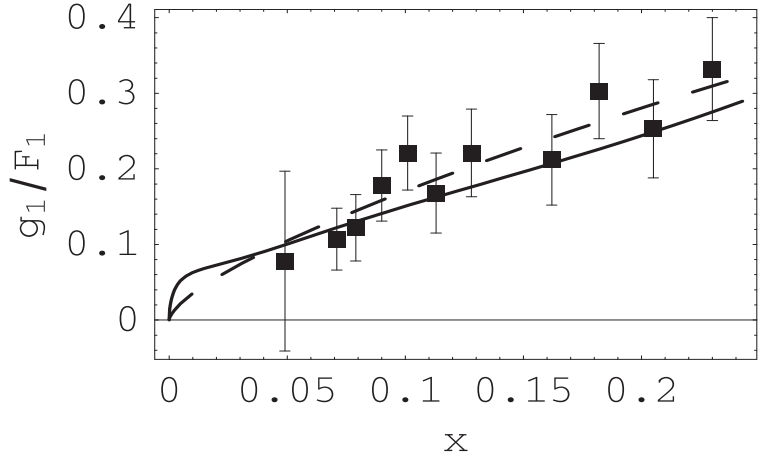,height=6.0cm,width=10.0cm}}
\vspace*{13pt}
\caption{Plots of ${\rm g_1(x,Q^2)/F_1(x,Q^2)}$ for ${\rm Q^2=1~GeV^2}$.
The solid and dashed curves are the solution of the DGLAP evolution equation
and the experimental fit correspondingly.
Experimental points with total errors are taken from the paper [21].}
\end{figure}

\begin{figure}[htbp]\vspace*{0.0cm}
\epsfxsize=0.9\textwidth
\centerline{\psfig{figure=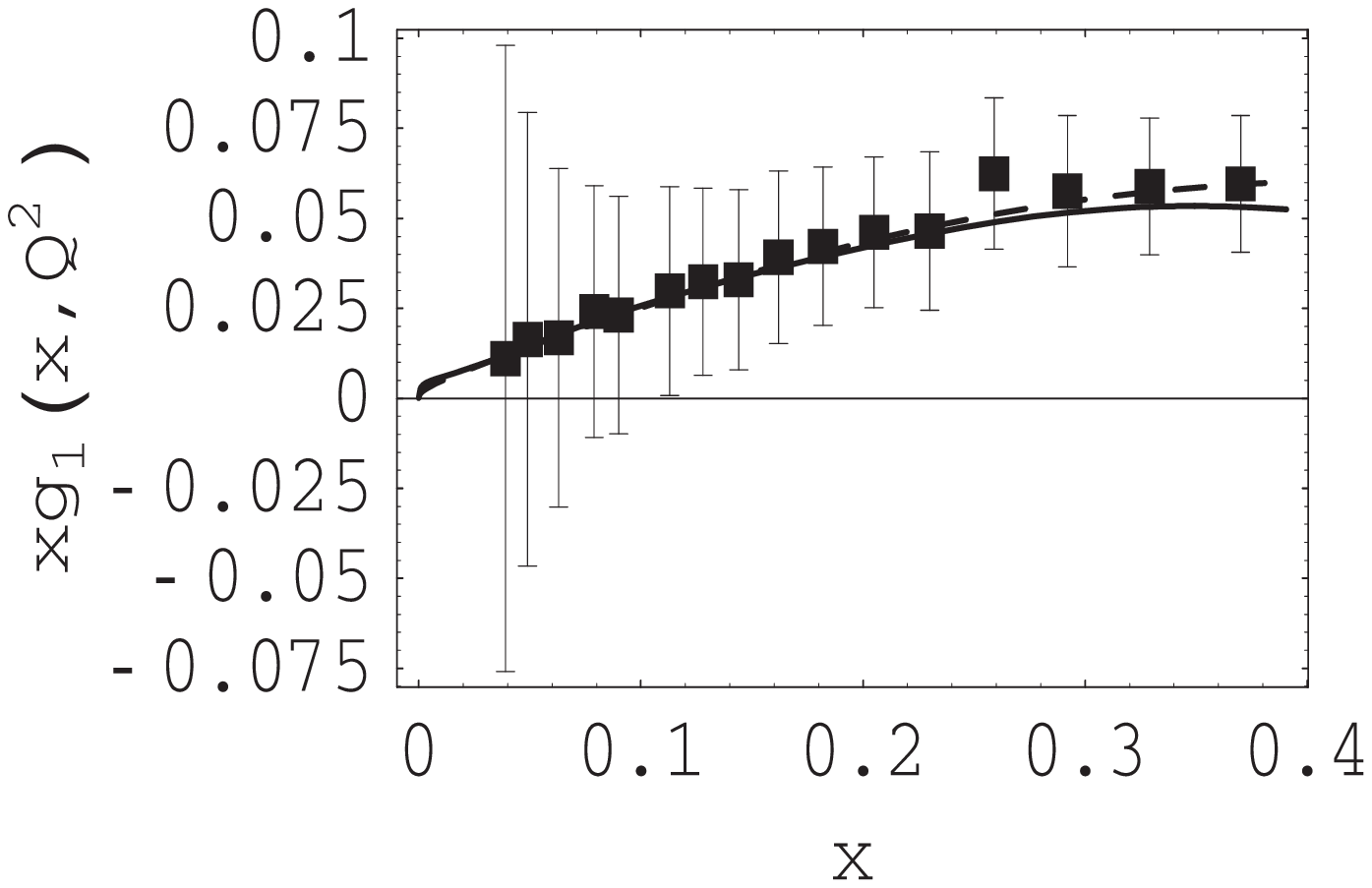,height=6.0cm,width=10.0cm}}
\vspace*{13pt}
\caption{Proton structure function ${\rm xg_1(x,Q^2)}$ for ${\rm Q^2=2~GeV^2}$ in
the nonresonance region (solid curve). The dashed curve is the experimental fit.
Experimental points correspond to the paper [21]}
\end{figure}

\begin{figure}[htbp]\vspace*{0.0cm}
\epsfxsize=0.9\textwidth
\centerline{\psfig{figure=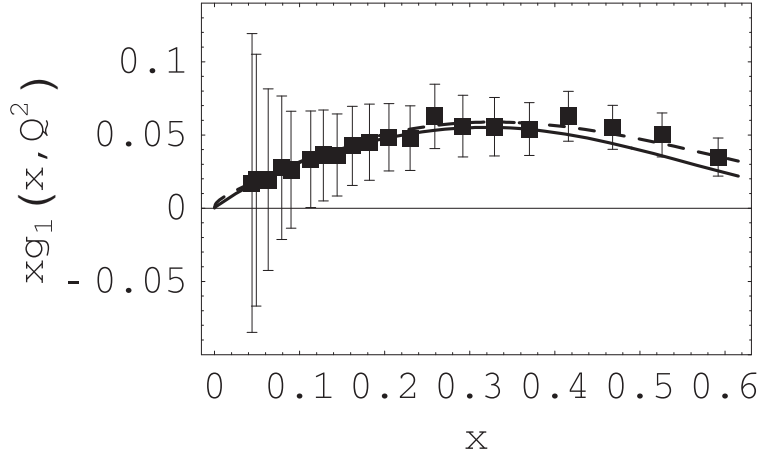,height=6.0cm,width=10.0cm}}
\vspace*{13pt}
\caption{Proton structure function ${\rm xg_1(x,Q^2)}$ for ${\rm Q^2=5~GeV^2}$
in the nonresonance region (solid curve). The dashed curve is the experimental fit.
Experimental points correspond to the paper [21].}
\end{figure}

\subsection{Resonance region}

The proton polarizability contribution to HFS in the resonance
region is determined by the processes of photo- and
electroproduction on nucleons of the pions and some prominent
baryon resonances. The amplitudes of such reactions are shown on
Fig.5.

To obtain correction (8) at the resonance region ${\rm (W^2\leq 4
GeV^2)}$ we use the Breit-Wigner parameterization for the
photoabsorption cross sections in (18), (19), suggested in
\cite{Walker,Arndt,Teis1,Teis2,Krusche,Bianchi,D}. There are many
baryon resonances that give contribution to photon absorption
cross sections. We take into account only five most important
resonances: ${\rm P_{33} (1232)}$, ${\rm S_{11} (1535)}$, ${\rm D_{13} (1520)}$,
${\rm P_{11} (1440)}$, ${\rm F_{15} (1680)}$. Considering the one-pion decay
channel of the resonances, the absorption cross sections ${\rm
\sigma_{1/2}}$ and ${\rm \sigma_{3/2}}$ may be written as follows
\cite{Teis2,Dong1}:
\begin{equation}
{\rm \sigma_{1/2,3/2}=\left(\frac{k_R}{k}\right)^2\frac{W^2\Gamma_\gamma\Gamma_{R
\rightarrow N\pi}}{(W^2-M_R^2)^2+W^2\Gamma_{tot}^2}\frac{4m_p}{M_R\Gamma_R}
|A_{1/2,3/2}|^2}
\end{equation}
where ${\rm A_{1/2,3/2}}$ are transverse electromagnetic helicity
amplitudes,
\begin{equation}
{\rm \Gamma_\gamma=\Gamma_R\left(\frac{k}{k_R}\right)^{j_1}\left(\frac{k_R^2+X^2}
{k^2+X^2}\right)^{j_2},~~X=0.3~{\rm GeV}.}
\end{equation}

\begin{figure*}[t!]
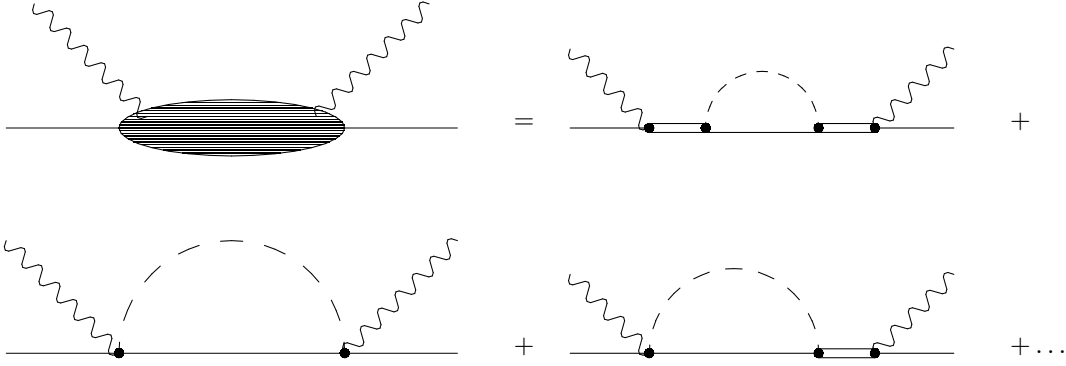

\magnitude=2000
\GRAPH(hsize=15){
\mov(0,0){\lin(1,0)}%
\mov(3,0){\lin(1,0)}%
\mov(0,-2){\lin(4,0)}%
\mov(1.,-2){\Circle*(0.08)}%
\mov(3.,-2){\Circle*(0.08)}%
\mov(2,-2){\halfdashcirc(2.)[U]}%
\mov(1.,-2){\wavelin(-1.0,1.)}%
\mov(3,-2){\wavelin(1.0,1.)}%
\mov(2.,0){\Ellipse*(0.5)[4,1]}%
\mov(1.25,0.1){\wavelin(-1.0,1.)}%
\mov(2.75,0.1){\wavelin(1.0,1.)}%
\mov(4.5,0){=}%
\mov(4.5,-2){+}%
\mov(5.0,0){\lin(0.7,0)}%
\mov(5.7,0){\Circle*(0.08)}%
\mov(5.7,0){\wavelin(-0.7,0.7)}%
\mov(5.7,-0.04){\lin(0.5,0)}%
\mov(5.7,0.04){\lin(0.5,0)}%
\mov(6.2,0){\Circle*(0.08)}%
\mov(6.2,-0.04){\lin(1,0)}%
\mov(7.7,0){\Circle*(0.08)}%
\mov(7.7,0){\wavelin(0.7,0.7)}%
\mov(7.2,-0.04){\lin(0.5,0)}%
\mov(7.2,0.04){\lin(0.5,0)}%
\mov(7.2,0){\Circle*(0.08)}%
\mov(7.7,0){\lin(0.7,0)}%
\mov(6.7,0){\halfdashcirc(1.0)[U]}%
\mov(8.9,0){+}%
\mov(8.9,-2){$+\ldots$}%
\mov(5.0,-2){\lin(0.7,0)}%
\mov(5.7,-2){\Circle*(0.08)}%
\mov(5.7,-2){\wavelin(-0.7,0.7)}%
\mov(5.7,-2){\lin(1.5,0)}%
\mov(7.7,-2){\Circle*(0.08)}%
\mov(7.7,-2){\wavelin(0.7,0.7)}%
\mov(7.2,-1.96){\lin(0.5,0)}%
\mov(7.2,-2.04){\lin(0.5,0)}%
\mov(7.2,-2){\Circle*(0.08)}%
\mov(7.7,-2){\lin(0.7,0)}%
\mov(6.45,-2){\halfdashcirc(1.5)[U]}%
}
\vspace*{13pt}
\caption{Feynman amplitudes for proton
polarizability correction in the resonance region. Solid, double
solid, wave and dashed lines correspond to nucleon, baryon
resonance, photon and pion correspondingly.}
\end{figure*}

The resonance parameters ${\rm \Gamma_R}$, ${\rm M_R}$, ${\rm j_1}$, ${\rm j_2}$,
${\rm \Gamma_{tot}}$
were taken from \cite{RPP,Teis3}. In accordance with Refs. \cite{Teis1,Krusche,Teis3}
the parameterization of one-pion decay width is
\begin{equation}
{\rm \Gamma_{R\rightarrow N\pi}(q)=\Gamma_R\frac{M_R}{M}\left(\frac{q}{q_R}\right)^3
\left(\frac{q_R^2+C^2}{q^2+C^2}\right)^2,~~C=0.3~{\rm GeV}}
\end{equation}
for the ${\rm P_{33}(1232)}$ and
\begin{equation}
{\rm \Gamma_{R\rightarrow N\pi}(q)=\Gamma_R\left(\frac{q}{q_R}\right)^{2l+1}
\left(\frac{q_R^2+\delta^2}{q^2+\delta^2}\right)^{l+1},}
\end{equation}
for ${\rm D_{13}(1520)}$, ${\rm P_{11}(1440)}$, ${\rm
F_{15}(1680)}$. ${\rm l}$ is the pion angular momentum and
$\delta^2$ = ${\rm (M_R-}$ ${\rm m_p-m_\pi)^2}$ + ${\rm \Gamma_R^2/4}$. Here q (k)
and ${\rm q_R}$ ${\rm (k_R)}$ denote the c.m.s. pion (photon)
momenta of resonances with mass M and ${\rm M_R}$ respectively.
In the case of $S_{11}(1535)$ we take into account ${\rm \pi N}$
and ${\rm \eta N}$ decay modes \cite{Krusche,Teis3}:
\begin{equation}
{\rm \Gamma_{R\rightarrow\pi,\eta}=\frac{q_{\pi,\eta}}{q}b_{\pi,\eta}\Gamma_R
\frac{q_{\pi\eta}^2+C_{\pi,\eta}^2}{q^2+C_{\pi,\eta}^2},}
\end{equation}
where ${\rm b_{\pi,\eta}}$ is the $\pi$ ($\eta$) branching ratio.

The cross section ${\rm \sigma_{TL}}$ is determined by an expression
similar to (25), containing product ${\rm (S^\ast_{1/2}\cdot
A_{1/2}+A_{1/2}^\ast S_{1/2})}$ \cite{Abe1}. The calculation of
helicity amplitudes ${\rm A_{1/2}}$, ${\rm A_{3/2}}$ and
longitudinal amplitude ${\rm S_{1/2}}$, as functions of ${\rm
Q^2}$, was done on the basis of constituent quark model (CQM) in
\cite{Dong2,Isgur,CL,Capstick,LBL,Warns}. In the real photon
limit ${\rm Q^2=0}$ we take corresponding resonance amplitudes from
\cite{RPP}. At low center of mass energies the excitation of
${\rm \Delta(1232)}$ resonance is of particular importance. At small
${\rm Q^2}$ it is dominated by a magnetic dipole transition.
The ${\rm Q^2}$ dependence of ${\rm A_{1/2}(Q^2)}$, ${\rm A_{3/2}(Q^2)}$
was extracted from the analysis made in \cite{Carlson}. Helicity
amplitudes of the other resonances were taken from the calculations
in CQM \cite{CL,Capstick,LBL,Warns}. We have considered Roper
resonance ${\rm P_{11}(1440)}$ as an ordinary ${\rm qqq}$ state
\cite{MAP}. As it follows from predictions of the quark model,
the helicity amplitudes, which may be suppressed at ${\rm Q^2=0}$,
become dominant very rapidly with increasing $Q^2$.

The two-pion decay modes of the higher nucleon resonances (${\rm
S_{11}(1535)}$, ${\rm D_{13}(1520)}$, ${\rm P_{11}(1440)}$ and
${\rm F_{15}(1680)}$ were described phenomenologically using
two-step process as in \cite{Teis1}. The high lying nucleon
resonance R can decay first into ${\rm N^{\ast}}$ (${\rm
P_{33}(1232)}$ or ${\rm P_{11}(1440)}$) and a pion or into a
nucleon and $\rho$- or $\sigma$-meson. Then the new resonances
decay into a nucleon and a pion or two pions:
\begin{displaymath}
{\rm R\rightarrow r+a=\Biggl\{{N^\ast+\pi\rightarrow
N+\pi+\pi,\atop \rho(\sigma)+N\rightarrow N+\pi+\pi.}}
\end{displaymath}
The total decay width of such processes can be presented as a
phase space weight integral over the mass distribution of the
intermediate resonance r= ${\rm N^\ast, \rho, \sigma}$ (${\rm
a=\pi, N}$):
\begin{equation}
{\rm \Gamma_{R\rightarrow
r+a}(W)=\frac{P_{2\pi}}{W}\int_0^{W-m_a}d\mu \cdot
p_f\frac{2}{\pi}
\frac{\mu^2\Gamma_{r,tot}(\mu)}{(\mu^2-m_r^2)^2+\mu^2\Gamma_{r,tot}^2(\mu)}
\frac{(M_R-m_2-2m_\pi)^2+C^2}{(W-m_2-2m_\pi)^2+C^2},}
\end{equation}
where ${\rm C=0.3~GeV}$, the factor ${\rm P_{2\pi}}$ must be
taken from the constraint condition: ${\rm \Gamma_{R\rightarrow
r+a}(W_R)}$ coincides with the experimental data, ${\rm p_f}$ is
the three momentum of resonance r in the rest frame of R.
${\rm \Gamma_{r,tot}}$ is the total width of the resonance r. The
decay width of the meson resonance is parameterized similarly to
that of the ${\rm P_{33}(1232)}$:
\begin{equation}
{\rm \Gamma(\mu)=\Gamma_r\frac{m_r}{\mu}\left(\frac{q}{q_r}\right)^{2J_r+1}
\frac{q_r^2+\delta^2}{q^2+\delta^2},~~~\delta=0.3~GeV,}
\end{equation}
where ${\rm m_r}$ and $\mu$ are the mean mass and the actual mass of the meson
resonance, q and ${\rm q_r}$ are the pion three momenta in the rest frame of the
resonance with masses $\mu$ and ${\rm m_r}$, ${\rm J_r}$ and ${\rm \Gamma_r}$
are the spin and decay width of the resonance with mass ${\rm m_r}$.

To calculate the nonresonant contributions to cross sections ${\rm
\sigma_{1/2}}$, ${\rm\sigma_{3/2}}$, ${\rm\sigma_{TL}}$ in the
resonance region we have used predictions of the unitary isobar
model (UIM) \cite{D,DKKPT,DKT}. Nonresonant background
contribution to ${\rm g_{1,2}(x,Q^2)}$ for ${\rm W^2\leq 4~GeV^2}$
can be derived from effective Lagrangians describing the
electromagnetic ${\rm\gamma N N}$, ${\rm \gamma\pi\pi}$
interactions and the hadronic ${\rm\pi NN}$ system \cite{D}. These
Lagrangians lead to the well-known expressions for the CGLN
amplitudes ${\rm F_1,\ldots , F_6}$ \cite{CGLN}, which give the
dominant part of the background. The other part is related to
vector meson exchange contributions. Within UIM taking into
account Born terms, vector meson terms and the interference
resonance background contributions to the polarized structure
functions we can calculate the single-pion production amplitudes
(see Fig.5) using the on-line version of the numerical program
MAID: http://www.kph-uni-mainz.de/MAID. We used it to consider
nonresonance pion electroproduction contributions to
$\Delta_{1,2}$.

The Breit - Wigner five resonance
parameterization of photon cross sections and constituent quark
model results give good description of proton polarized structure
functions in the resonance region. But the still existing difference
between this description of ${\rm g_{1,2}(\nu,Q^2)}$ and experimental data
\cite{FM2000,Dong1} requires further improvement
in the construction of spin dependent structure functions. It can
be done considering contributions of additional baryon
resonances in the large W range: ${\rm S_{31}(1620)}$, ${\rm F_{37}(1950)}$,
${\rm D_{33}(1700)}$, ${\rm P_{13}(1720)}$, ${\rm F_{35}(1905)}$ and accounting
for different decay modes of such states. The particular significance in the
study of the spin-dependent properties of baryon resonances belongs
to Gerasimov-Drell-Hearn (GDH) sum rule \cite{GDH}
\begin{equation}
{\rm -\frac{\kappa^2}{4m_2^2}=\frac{1}{8\pi^2\alpha}\int_{\nu_{th}}^\infty
\frac{d\nu}{\nu}[\sigma_{1/2}(\nu,0)-\sigma_{3/2}(\nu,0)]}
\end{equation}
The GDH sum rule rests on the basic physical principles and an unsubtracted
dispersion relation applied to the forward Compton amplitude. Recently
there was obtained first experimental data for the contribution to GDH sum
rule in the energy range ${\rm 200\leq E_\gamma\leq 800~MeV}$ \cite{Ahrens}.
Our expressions for the polarized structure functions ${\rm g_{1,2}(x,Q^2)}$
show that sum rule (32) is valid with high accuracy. Here we must call special
attention to the fact that the most important region for the integrals (9) and
(10) on ${\rm Q^2}$ variable lies from ${\rm 0.1~ Gev^2}$ to ${\rm 1~GeV^2}$.

The second part of (9) gives especially large negative contribution to the
correction ${\rm \delta_1^P}$ in the range of small $Q^2$, where the
contribution of $\Delta$ isobar is dominant. With increasing
$Q^2$ its value falls and the total correction ${\rm \delta_1^P}$
has positive sign.

\section{Numerical results}

The processes of electromagnetic production of light quark baryon
resonances represent the unique sphere for the investigation of
nonperturbative properties of quantum chromodynamics (QCD). The direct
test of the hadron quark model, chiral perturbation theory (ChPT), UIM,
QCD sum rules and other approaches is connected with the study of the GDH
sum rule \cite{Ahrens},
helicity amplitudes ${\rm A_{1/2}}$, ${\rm A_{3/2}}$, ${\rm S_{1/2}}$
and transition form factors \cite{Frolov,Stoler}, the nucleon
spin-dependent structure functions \cite{Abe1,Abe2,Anthony}. The HFS
measurement in the hydrogen atom and in the muonic hydrogen can be
considered also for the verification of different effective field theories,
because ${\rm \Delta E_{HFS}(ep)}$ is one of the most precisely measured
quantities in atomic physics.

In this work for the ground state hyperfine splitting interval we
considered the part of the strong interaction effects, which manifest
itself in the proton polarizability correction. The calculation of ${\rm\delta^P}$
was based on three main ingredients:\\
1. The evolution equations
for the spin dependent structure functions in the nonresonance region,\\
2. The phenomenological hadron quark model, isobar model
in the resonance region.\\
3. The experimental data for the nucleon polarized structure functions
obtained at SLAC, DESY, CERN.

The values of contributions $\delta_1^P$, $\delta_2^P$ and the total
contribution $\delta^P$, obtained after the numerical
integration in the resonance and nonresonance regions are as
follows:
\begin{equation}
{\rm \delta_{1,res}^P=3.13\cdot 10^{-4},~~~\delta_{1,nonres}^P=
2.01\cdot 10^{-4},~~\delta_1^P=5.14\cdot 10^{-4},}
\end{equation}

\begin{equation}
{\rm \delta_{2,res}^P=-0.56\cdot 10^{-4},~~~\delta_{2,nonres}^P=
-0.02\cdot 10^{-4},~~\delta_2^P=-0.58\cdot 10^{-4},}
\end{equation}

\begin{equation}
{\rm \delta^P=\delta_1^P+\delta_2^P=4.6\pm 0.8\cdot 10^{-4},}
\end{equation}
where the error, indicated in the expression (35), is determined
by three main factors, connected with the polarized structure
functions. We solved DGLAP equations in the NLO approximation, so
possible uncertainty in ${\rm \delta^P}$ can comprise near 10$\%$
of obtained result. The other source of the theoretical
uncertainty arises from the experimental data errors in the ${\rm
Q^2\le 1~GeV^2}$ region. We estimated it at a level of about 20
$\%$ of the contribution ${\rm \delta^P}$ at ${\rm Q^2\le
1~GeV^2}$ in the nonresonance region. Finally, the third part of
uncertainty in (35) is controlled by the predictions of the
constituent quark model for the electroproduction amplitudes
${\rm A_{1/2}}$, ${\rm A_{3/2}}$, ${\rm S_{1/2}}$ \cite{B}. We
supposed that this one can reach 20 $\%$ of the contribution
$\delta^P$ in the resonance region, comparing expressions
(18)-(19) with the experimental data in the resonance region
\cite{FM2000}. On our opinion this is the main source of
theoretical uncertainty now. The first way to obtain more
reliable result in the resonance region can be based on using
transition form factors of the nucleon to the baryon resonances
with different values of ${\rm J^P}$, calculated on the basis of
QCD sum rules as in \cite{FMS} and of ChPT approach. The second
one is connected with the progress in measurement of polarized
structure functions ${\rm g_{1,2}(x,Q^2)}$ and perhaps lattice
calculations.

Muon mass dependent terms in relations (8)-(10), which are
negligible for electronic hydrogen, give contribution to
$\delta^P$ at a level of 25 $\%$. In the case of muonic hydrogen
${\rm E^F(\mu p)=182.443~ meV}$ and the value of proton
polarizability contribution to the HFS is equal to 0.084 meV for
n=1 and 0.011 meV for n=2 state. Last value can be important for
the determination of the Lamb shift from the 2P-2S frequency
measurement \cite{FK}. The total correction (35) must be taken
into account if the measurement of the ground state HFS in muonic
hydrogen with the accuracy $10^{-4}$ became available \cite{BR}.\\[2mm]

\begin{acknowledgements}

We are grateful to M.Hirai, S.Kumano and M.Miyama for sending us the Fortran
program for solving DGLAP evolution equations and to D.Bakalov, F.Kottman,
V.Savrin for many useful discussions.
The work was performed under the financial support of the Russian Foundation
for Fundamental Research (grant 00-02-17771), the Program "Universities
of Russia - Fundamental Researches" (grant 990192) and the Ministry of Education
(grant EOO-3.3-45).
\end{acknowledgements}

\end{document}